\documentclass[preprint]{aastex}
\usepackage{emulateapj5}
\usepackage{apjfonts}

\newcommand{\gtilde}
 {~ \raisebox{-1ex}{$\stackrel{\textstyle >}{\sim}$} ~}
\newcommand{\ltilde}
 {~ \raisebox{-1ex}{$\stackrel{\textstyle <}{\sim}$} ~}
\def\ltsima{$\; \buildrel < \over \sim \;$}
\def\ltsim{\lower.5ex\hbox{\ltsima}}
\def\gtsima{$\; \buildrel > \over \sim \;$}
\def\gtsim{\lower.5ex\hbox{\gtsima}}

\submitted{To Appear in the Astrophysical Journal Letters}

\shorttitle{HEROS IN THE SUBARU DEEP FIELD}
\shortauthors{TOTANI ET AL.}

\begin{document}

\title{Hyper Extremely Red Objects in the Subaru Deep Field:
Evidence for Primordial Elliptical Galaxies in the Dusty Starburst Phase$^*$
}

\author{Tomonori Totani$^{2, 8}$, Yuzuru Yoshii$^{3, 4}$,
Fumihide Iwamuro$^5$, Toshinori Maihara$^6$, and Kentaro Motohara$^7$}

\altaffiltext{2}{Theory Division, 
National Astronomical Observatory, Mitaka, Tokyo 181-8588,
Japan (E-mail: totani@th.nao.ac.jp)}
\altaffiltext{3}{Institute of Astronomy, School of Science,
The University of Tokyo, 2-21-1 Osawa, Mitaka, Tokyo 181-8588, Japan}
\altaffiltext{4}{Research Center for the Early Universe, School of Science,
The University of Tokyo, Tokyo 113-0033, Japan}
\altaffiltext{5}{Department of Physics, Kyoto University, Kitashirakawa,
Kyoto 606-8502, Japan}
\altaffiltext{6}{Department of Astronomy, Kyoto University, Kitashirakawa,
Kyoto 606-8502, Japan}
\altaffiltext{7}{Subaru Telescope, National Astronomical Observatory of Japan,
650 North A'ohoku Place, Hilo, HI 96720, USA}
\altaffiltext{8}{
Present Address: Princeton University Observatory, Peyton Hall,
Princeton, NJ 08544-1001, USA}

\begin{abstract}
We report observational analyses and theoretical interpretations of
unusually red galaxies in the Subaru Deep Field (SDF). 
A careful analysis of the SDF data
revealed a population with unusually red near-infrared (NIR)
colors of $J - K \gtilde$ 3--4, with higher confidence than the
previous SDF result. Their surface
number density drastically increases at $K \gtilde 22$ and becomes
roughly the same with that of dusty starburst galaxies detected by
submillimeter observations in recent years. These colors are even redder than
the known population of the extremely red objects (EROs), and too red
to explain by passively evolving elliptical galaxies which are the
largest population of EROs.
Hence these hyper extremely red objects (HEROs)
should be considered as a distinct population from EROs. 
We discuss several possible interpretations of these enigmatic objects, and 
we show that these red NIR colors, $K$-band and
sub-mm flux, and surface number density are quantitatively best explained by
primordial elliptical galaxies reddened by dust, still in the starburst phase
of their formation at $z \sim 3$. 
\end{abstract}


\noindent
$^*$Based on the data corrected at the Subaru telescope, which is operated
by the National Astronomical Observatory of Japan.

\section{Introduction}
The reddest populations of galaxies are of great interest
in the study of galaxy formation and evolution, since they are
showing one of the most extreme aspects in the history of galaxies
and structure formation. A population called extremely red objects
(EROs) is already established, showing very red colors between
the optical and NIR bands, e.g., $R - K \gtilde$ 5--6
(e.g., Elston, Rieke \& Rieke 1988; McCarthy, Perrson, \& West 1992; 
Hu \& Ridgway 1994;
Thompson et al. 1999; Yan et al. 2000; Scodeggio \& Silva 2000;
Daddi et al. 2000a,b). Daddi et al. (2000a) argued that
a large part of this population consists of passively evolving 
elliptical galaxies at $z \sim$ 1--2, based on the strong clustering
of field EROs. A smaller fraction of EROs seems to be
ultra-luminous infrared
galaxies (ULIRGs) with $L_{\rm IR} > 10^{12} L_\odot$ under dusty
starbursts, some of which are also observed as
submillimeter sources (Cimatti et al. 1998; Dey et al. 1999;
Smail et al. 1999; Gear et al. 2000).

The process by which elliptical galaxies form is an important issue in galaxy
formation and cosmology.  One popular scenario is that they are formed with an
intense initial starburst, followed by passive evolution without star formation
to the present (Larson 1974; Arimoto \& Yoshii 1987). 
Passively evolving elliptical galaxies
are expected to show very red optical-NIR 
colors at $z \sim$ 1--2. The colors and
number density of EROs are well consistent with this picture of
elliptical galaxy formation (Daddi, Cimatti, \& Renzini 2000b).
On the other hand, the starburst phase at the formation of elliptical
galaxies has not yet been discovered. If the starburst phase is not
dusty their redshifted UV radiation should have easily been detected by the
past optical surveys (e.g., Totani, Sato, \& Yoshii 1997), 
and hence it has been suggested that elliptical
galaxies have formed with dusty starbursts or hierarchical merging of
smaller objects at relatively low redshifts (e.g., Zepf 1997).

Therefore search for dusty 
starbursts at high redshift is very important to verify the above
picture of elliptical galaxy formation. In recent years, there has been
dramatic progress in submillimeter observations, and it 
has revealed redshifted dust emission from high-$z$ starbursts 
(Hughes et al. 1998; Barger et al. 1998).  
Although their inferred star formation rates 
($\gg 100 M_\odot$/yr) are consistent
with those expected in the ``monolithic-like'' collapse scenarios,
the evidence for connection between these SCUBA sources and elliptical
galaxies is rather weak based on the submillimeter observations alone,
due to the lack of a quantitative connection with
present-day elliptical galaxies. In fact, optical observations have failed to
identify the counterparts for most of the SCUBA sources (Dunlop 2001).  

Here we report a discovery of unusual objects 
in the Subaru Deep Field (SDF, Maihara et al. 2001), 
which are even redder than EROs and too red to explain by passively evolving
elliptical galaxies at $z \lesssim 2$. 
Therefore these hyper extremely red objects (HEROs)
should be considered as a distinct population from previously known EROs.
The SDF is one of the deepest images of the universe yet obtained in
the NIR bands. The $J$ and $K$ band images have been
taken for a $2'\times 2'$ field, 
with the limiting magnitudes of $J$  = 25.1 and $K$ = 23.5 at 5$\sigma$ level. 
SDF found 385 and 350 galaxies down to these magnitudes, respectively.
The first-pass analysis of this image revealed
four very red objects in the $J-K$ color. 
(The reddest color is $J-K = 4.12 \pm
1.04$ for an object with $K = 22.31 \pm 0.14$, see Maihara et al. 2001.) 
However, because of possible observational uncertainties in
the photometry, it was not strongly claimed that there really exist objects
with $J - K > 4$.  What we report here is a new, more careful analysis, to
investigate how many objects with such red colors actually exist in the SDF.
Indeed, we obtained stronger evidence for the existence of such unusually
red NIR color objects with $J-K>$ 3--4.  
We will then discuss several possible interpretations of these HEROs, and
argue that these objects are likely to be primordial elliptical galaxies
in the dusty starburst phase. 

\section{HEROs in SDF}
\begin{figure*}
\plotone{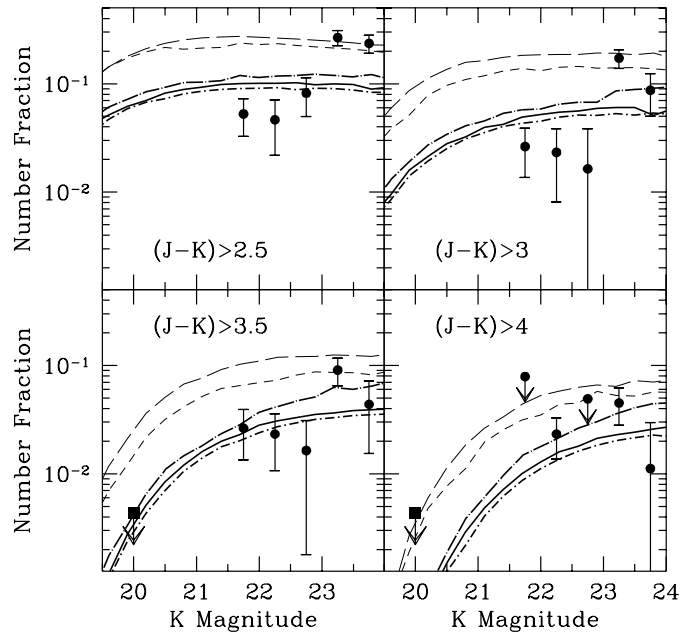}
\caption{Number fraction of galaxies redder than several threshold $J-K$ colors
(indicated in each panel), as a function of $K$ magnitude.  Filled circles are
the data of the SDF.  
The error bars are 1$\sigma$, while the upper limits shown by arrows
are at the 95\% confidence level. 
The upper limit at $K = 20$ is from Scodeggio \& Silva (2000, 
filled square).  The
solid line is the model prediction with the formation redshift $z_F = 3$ and
our standard dust-extinction normalization.  The short- and long-dot-dashed
lines are the same as the solid line, but for the cases with dust-extinction
normalization multiplied by factors of 2 and 1/2, respectively.  The thin
short- and long-dashed lines are the same as the solid line, but for $z_F = 5$
and 7, respectively.  
}
\label{fig:mag-frac}
\end{figure*}

We estimated how many objects really exist in the SDF, with colors redder than
a given $J-K$ color. 
The results are shown by a number fraction of such red objects as a function
of $K$ magnitudes in Fig. \ref{fig:mag-frac}, down to magnitudes
fainter than the previously reported four objects. 
We should be very careful in this analysis, because
the $J$ band flux of red objects is very faint by definition, 
although they may be detected easily in the $K$-band.
We took the analysis procedure as follows to carefully take into account the
effects of photometric uncertainty.
First, SDF objects are distributed onto two-dimensional bins in $K$
magnitude and $J-K$ color.  Expected counts of spurious objects are then
subtracted from each bin, based on their estimated color distribution from the
reference frame (Maihara et al. 2001). 
(In fact, since the 5$\sigma$ detection limit of the SDF is $K=23.5$, 
contamination of spurious objects is not significant at $K < 23.5$.)
To estimate the effect of photometric
uncertainties, we performed Monte Carlo simulations in which the magnitudes and
colors are perturbed according to their estimated error distributions; the
resulting dispersion of the counts in each bin is added to the error arising
from spurious object subtraction.  Here, an asymmetric distribution of photon
counts is used for the photometric error distribution, 
also taking into account the
galaxy count slope (see Maihara et al. 2001 for details).  
This final error estimate is then shown in Fig. \ref{fig:mag-frac}.

The data show that there exists a very red population with $(J - K) >$ 3--4,
having a number fraction of 1--10\% at $K \gtilde 22$, about one order of
magnitude higher than that at brighter magnitudes of $K \ltilde 20$
(Scodeggio \& Silva 2000). 
Dickinson et al. (2000) has reported an unusual object with $K = 22.0$ and
having a very red color $(J-K > 4.6)$ in the Hubble Deep Field (HDF). This
object is most likely a brighter example of the population we have
found here. Saracco et al. (2001) found a fraction ($\sim 5$\%) of
sources with color redder than $J-K_s=2.3$ at magnitudes $K_s > 20$,
which is also consistent with our result.
The numbers of real objects estimated for the
reddest galaxies in the SDF with $(J-K)>4$ are $1.0 \pm 0.41 (2.4\sigma)$ and
$3.81 \pm 1.44 (2.6\sigma)$ in $K$ = 22.0--22.5 and 23.0--23.5, respectively.

Even EROs do not show unusually red colors within the NIR
bands; typical EROs have colors of $J-K \sim$ 2 (Scodeggio \& Silva
2000).  These colors should also be compared with typical theoretical 
predictions for passively evolving elliptical galaxies. 
Here we use a typical model of elliptical galaxies 
($M_B = -20$) without dust obscuration (Kodama \& Arimoto 1997). A low-density
flat universe with $(h, \Omega_0,\Omega_\Lambda)$ = (0.7, 0.2, 0.8) is
assumed here and throughout this letter.
This model predicts
that the $J-K$ color should be bluer than 2.5 at $z \ltilde 2$ 
for any formation redshift (dotted lines in Fig.
\ref{fig:E-model}). This is well consistent with Fig. 1 of Scodeggio \&
Silva (2000) using a different evolution model of Bruzual \& Charlot
(1993). 

\section{Interpretations of HEROs}
\begin{figure*}
\plotone{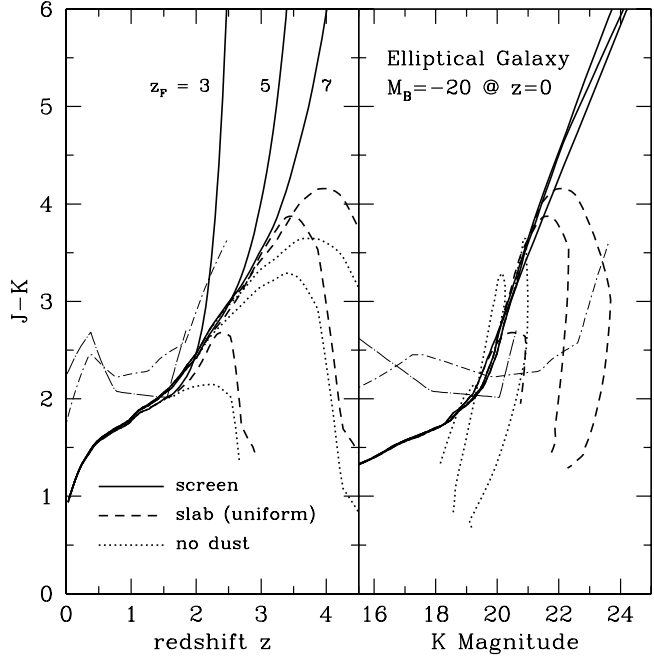}
\caption{
The $J-K$ color versus redshift and $K$ magnitude relation for a standard model
of elliptical galaxies 
with the present-day magnitude of $M_B = -20$, with different models
of dust extinction: no dust (dotted), slab-type dust (dashed), and
screen (solid line). 
Three curves are depicted for each of the three line
markings, corresponding to different values of formation redshift, $z_F =$ 3,
5, and 7, from left to right with increasing $z_F$.  
In addition, the colors and magnitudes of typical
``template'' starburst galaxies of Arp 220 (long-dot-dashed;
photometric data from Sanders et al. 1988 and Carico et al. 1990)
and the nuclear region of M82 (short-dot-dashed; photometric data
from Johnson 1966 and Ichikawa et al. 1995).
}
\label{fig:E-model}
\end{figure*}

Passively evolving ellipticals at even higher redshift (formation redshift
$z_F \gg 3$) may have redder colors, comparable with HEROs, but still the
maximum color does not exceed $J - K>4$ without reddening by dust (dotted lines
in Fig. \ref{fig:E-model}). Zepf (1997) argued that 
the absence of extremely red objects with $V_{606} - K > 7$ in HDF
rules out models in which typicall elliptical galaxies are fully
assembled and have formed all of their stars at $z \gtsim 5$.
Such a large $z_F$ is also disfavored from the
constraint that major episodes of star formation should have taken place at $z
\ltilde 3$, as inferred from the observed colors of elliptical galaxies
at $z \sim$ 1 (Franceschini et al. 1998). 
Two possibilities then remain to account for the origin
of objects with such red colors: (1) 
starburst galaxies obscured by dust, or (2) very
high-$z$ Lyman-break galaxies at $z \gtilde 10$. 
The possibilities of any point sources such as quasars
or very red stars in the Galaxy
are rejected by the extended profiles of the HERO images.
The exclusion of point sources in the SDF is secure down to
$K \sim 23$ (Nakajima et al. 2000), and 
see Maihara et al. (2001) for images of the four brightest HEROs.
We will then argue in the following
that the scenario (1) provides the best explanation for
presence of HEROs in the SDF data.

It is easy to show that elliptical galaxies 
at the formation stage are likely to be
very dusty, with an optical depth more than 10 for UV radiation, based on the
expected column density of dust (e.g., Totani \& Yoshii 2000).  
This is in contrast to spiral
disks, in which dust opacity does not become much higher than in present-day
galaxies because of the modest star formation generally believed to have taken
place (Totani \& Kobayashi 1999). 
We show the model colors of a typical elliptical galaxy in the presence
of dust obscuration in Fig. \ref{fig:E-model}, according to the dust modeling
used in Totani \& Yoshii (2000). The dust opacity is
assumed to be proportional to the gas column density and gas metallicity.  A
natural normalization factor for the dust opacity has been chosen to reproduce
a typical mean extinction of $A_V \sim 0.2$ for our Galaxy. The Galactic
extinction curve (e.g., Mathis, Mezger, \& Panagia 1983)
is used. It should be noted that we will mainly consider dusty galaxies at
$z \sim 3$, corresponding to the restframe wavelength of $\lambda > 3000$
{\AA} where the difference of extinction curves between the Galaxy,
Magellanic Clouds, and starburst galaxies is almost negligible
(e.g., Calzetti, Kinney, \& Storchi-Bergmann 1994).
In the model we
assumed that the gas fraction exponentially decays as $f_g \propto
\exp(-t/t_{\rm GW})$ after the time of galactic wind ($t_{\rm GW}
=0.3$Gyr here).  
The model prediction depends on the assumed spatial
distributions of dust: screen or slab (the same distribution for stars and
dust).  

Figure \ref{fig:E-model} shows that a typical giant elliptical galaxy with
screen-type dust shows a very red color of $(J-K) > 3.5$ at $K \gtilde 21$, in
good agreement with the properties of HEROs. The screen model is not
unreasonable, because the galactic wind, expected to be driven by the
strong starbursts of forming elliptical galaxies (Arimoto \& Yoshii
1987),  would also blow out
the dust particles. It also seems that at least a fraction of dust has a
screen-like distribution in observed starburst galaxies, as suggested by strong
reddening that cannot be explained only by the slab-type dust
model (Calzetti, Kinney, \& Storchi-Bergmann 1994;
Gordon, Calzetti, \& Witt 1997). The $J-K$ colors of typical
``template'' dusty starburst galaxies, i.e., Arp 220 and the nucleus of M82,
are also shown when they are placed at high redshifts. They are redder
than passively evolving elliptical galaxies at fixed redshifts, and
the color of M82 reaches $J-K \sim$ 3.5 at $z \sim 2.5$.

Therefore primordial elliptical galaxies in the starburst phase, obscured by
screen-like dust, are a promising candidate for the HEROs in the SDF.  To test
this hypothesis, we predict the expected surface number density of such objects
based on the number density of the present-day elliptical galaxies determined
by the local luminosity function (model curves in Fig.
\ref{fig:mag-frac}).  The model calculation is the same as
presented in Totani \& Yoshii (2000). 
The total counts of all galaxy types of this
model already fit well to the SDF counts
(Totani et al. 2001a, b), and hence agreement in
the number fraction simultaneously warrants the agreement in real counts.  The
model with $z_F = 3$, which is consistent with the redshift constraints of
Zepf (1997) and Franceschini et al. (1998), 
agrees well with the
observed number fraction of HEROs with $J-K \gtilde 3$--4 as a function of $K$
magnitude, lending further support for the hypothesis that HEROs in the SDF are
dusty primordial elliptical galaxies. 
The predictions with $z_F > 5$ overproduce HEROs
beyond the upper limit at $K=20$, while they are rather favored from the data
with $J - K> 2.5$ and $K > 23$, possibly due to the contribution of relatively
small, faint elliptical galaxies with large $z_F$.  This may suggest that
smaller elliptical galaxies have formed earlier than massive ones, as expected
from hierarchical structure formation.  To estimate the uncertainty concerning
dust extinction, we repeated the calculation with different values of the
normalization of dust optical depth. 
Changing this parameter by a factor of up to 4 does not
significantly change the prediction, as shown in the figure.

If, on the other hand, HEROs are high-$z$ Lyman-break galaxies, their redshift
must be $z \gtilde 10$ for the $J$ band to correspond to the restframe
wavelength of $\ltilde 1000$\AA. The absolute UV luminosity inferred from
$K\sim22$ then implies a star formation rate of $\gtilde 140 M_\odot$/yr,
ignoring dust extinction (see
Madau, Pozzetti, \& Dickinson 1998 for conversion factors).  
It would be difficult to convert all
baryonic gas within a time scale shorter than a typical dynamical time or
duration of starburst, i.e., $\sim 10^8$yr, and 
hence a conservative lower bound
for the baryonic mass of these systems is obtained as $M_B \gtilde 10^{10}
M_\odot$, or the total mass of $M \gtilde 10^{11} M_\odot$ including dark
matter.  We should then examine how many such massive objects could have
already formed at such high redshift, from the viewpoint of the standard theory
of bottom-up structure formation induced by cold dark matter (CDM).  We have
calculated the surface density of objects with $M > 10^{11} M_\odot$ at $z >
10$ using the standard Press-Schechter formalism in the 
$\Lambda$-CDM universe with $\sigma_8 = 1$, presently favored by various
cosmological observations, and found that it is about $\sim 0.16$ in the SDF.
[See, e.g., Kitayama \& Suto (1996) for the calculation method].  This is
considerably smaller than the observed number of HEROs in the SDF ($\sim 10$),
in spite of the conservative lower mass limit.  By contrast, the same
estimate for objects with $M \sim 10^{12-13} M_\odot$ at $z \sim$ 3--5,
expected typical values for dusty primordial elliptical galaxies, becomes about
30 in the SDF; this is then sufficient to explain the observed number of HEROs.
It should also be noted that very high-$z$ ($z > 5$) passively
evolving elliptical galaxies are again disfavored as the origin of
HEROs from this viewpoint.

\section{Possible Connection to Submillimeter Sources}
If HEROs are actually strong starbursts obscured by dust at $z \sim 3$, they
are expected to be strong sources in the submillimeter bands as a result of
their redshifted strong dust emission.  
SCUBA has revealed high-$z$ starbursts with surface density of about
$10^7 \ \rm sr^{-1}$ down to $S_\nu \sim$ 1 mJy in the 850$\mu$m
band (Hughes et al. 1998; Barger et al. 1998).  In the
following  we show that a considerable part of the SCUBA sources are
likely to have the same origin with HEROs in the SDF, for which we have already
shown a quantitative connection to the present-day elliptical galaxies.

First, the total galaxy counts (Maihara et al. 2001) 
down to $K \sim 22$--23 is about $10^5
\ \rm deg^{-2}$.  By using a fraction of about 3\% of the HEROs in the
SDF, the surface density of HEROs becomes $\sim 10^7 \ \rm sr^{-1}$, in good
agreement with the number density of SCUBA sources. Furthermore, we can predict
the submm flux from HEROs, based on the model used here.
We calculated the
total luminosity of dust emission from the amount of stellar light absorbed by
dust, and then obtained the expected mid- and far-infrared spectral
energy distribution (SED) of dust emission by a model (Takeuchi et al. 2001) 
based on the empirical relation between
total luminosity of dust emission and dust SED of local infrared
galaxies (e.g., Soifer \& Neugebauer 1991).  
The prediction for the observed flux and
SED for several values of redshifts is shown in Fig.
\ref{fig:fir} for a typical giant elliptical galaxy.  This figure shows that
dust emission from HEROs at $z \sim 3$ can actually be detected in the
850$\mu$m band with the SCUBA sensitivity ($S_\nu \sim$ mJy, shown by a filled
diamond at 850$\mu$m).  The redshift of $z \sim$ 3 inferred from our analysis
and Franceschini et al. (1998) is also well consistent with the median of the
redshift estimates for SCUBA sources (Dunlop 2001).

\section{Conclusions}
\begin{figure*}
\plotone{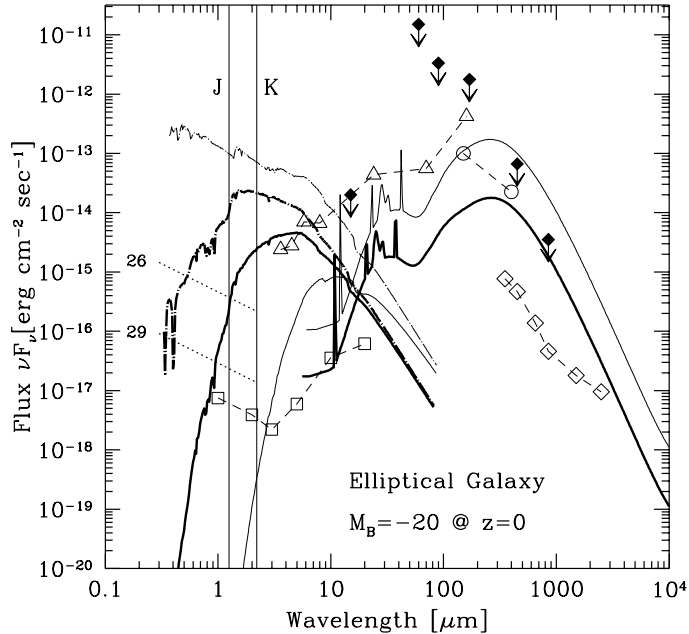}
\caption{
Spectral energy distribution (SED) of a typical elliptical galaxy with
present-day absolute magnitude of $M_B = -20$ and $z_F = 3$.  
The two components of direct stellar light surviving absorption 
(in the optical-NIR
band) and emission from heated dust (in the mid- to far-infrared band)
are shown by solid lines.  The
thick lines are for a galaxy at $z = 2.3$, at which its $J-K$ color becomes
4, while the thin lines are for a galaxy placed at $z = 2.7$, corresponding to
the epoch of the galactic wind.  The dot-dashed lines are the SED of direct
stellar light when there is no extinction by dust (again
thick and thin lines for $z
= $ 2.3 and 2.7, respectively).  Dotted lines in the optical and NIR
wavelengths show
the flux corresponding to the $AB$ magnitudes of 26 and 29, showing the
sensitivity limits of the SDF in the $K$ band and HDF in optical bands, 
respectively ($K_{AB} = K + 1.85$).  
The wavelengths corresponding to $J$ and $K$ bands are
indicated.  Filled diamonds with lower arrows are the sensitivity limits which
have been achieved by past instruments in infrared bands. 
The open symbols connected by dashed
lines are the expected sensitivities of future experiments: the
Herschel Space Observatory (circles),
roughly the same sensitivities of SIRTF and ASTRO-F (triangles), 
NGST (squares), and ALMA (diamonds). }
\label{fig:fir}
\end{figure*}

Based on the arguments presented above, we 
conclude that the best interpretation of the newly discovered
HEROs is that we are now beginning to detect high-$z$ dusty starbursts, whose
number density, NIR colors and magnitudes, 
and SED from NIR to submillimeter bands
are quantitatively consistent with the expectations of primordial
elliptical galaxies forming at $z \sim 3$.
An urgent task in the near future
is to establish the link between HEROs and SCUBA sources by 
follow-up observations for each population.
Further confirmation of our results would come in the more distant
future by next generation instruments, such as 
SIRTF, ASTRO-F, Herschel Space Observatory, 
NGST, and ALMA (see Fig. \ref{fig:fir} for their target sensitivities).
These will provide further valuable
information for the epoch of rapid and hidden
star formation in the early universe. 

We would like to thank T.T. Takeuchi for providing us with his infrared SED
model, and T.C. Beers for careful reading of this manuscript.


\epsscale{0.5}

\epsscale{0.5}

\end{document}